\documentclass[prb,showpacs,amsfonts,superscriptaddress,twocolumn]{revtex4}
\usepackage{epsfig}
\usepackage{dcolumn}
\usepackage{bm}
\usepackage{amsmath}
\usepackage{amsfonts}
\usepackage{amssymb}
\usepackage{graphicx}%
\begin{document}
\title{Scaling behavior of an Anderson impurity close to the Mott-Anderson
transition}
\author{M. C. O. Aguiar}
\affiliation{Center for Materials Theory, Serin Physics Laboratory, Rutgers
University, 136 Frelinghuysen Road, Piscataway, New Jersey 08854}
\author{V. Dobrosavljevi\'{c}}
\affiliation{Department of Physics and National High Magnetic Field
Laboratory, Florida State University, Tallahassee, FL 32306}
\author{E. Abrahams}
\affiliation{Center for Materials Theory, Serin Physics Laboratory, Rutgers
University, 136 Frelinghuysen Road, Piscataway, New Jersey 08854}
\author{G. Kotliar}
\affiliation{Center for Materials Theory, Serin Physics Laboratory, Rutgers
University, 136 Frelinghuysen Road, Piscataway, New Jersey 08854}


\begin{abstract}
The scaling behavior of Anderson impurity models each of which with a
different site
energy $\varepsilon_{i}$ is examined close to the Mott-Anderson transition.
Depending on its impurity energy $\varepsilon_{i}$, in the critical regime a
site turns into a local magnetic moment, as indicated by a vanishing
quasiparticle weight $Z\rightarrow0$, or remains nearly doubly occupied or
nearly empty, corresponding to $Z\rightarrow1$. In this paper, we present the
scaling behavior of $Z$ as a function of the on-site energy $\varepsilon_{i}$
and the distance $t$ to the transition, and interpret our result in terms of
an appropriate beta function $\beta(Z)$.
\end{abstract}
\maketitle

\section{Introduction}

The Mott~\cite{mott74} and the Anderson~\cite{anderson58} routes to
localization have long been recognized as two of the basic processes that
can drive the metal-insulator transition. Theories separately describing
each of these mechanisms were discussed long ago. According to the scaling
theory of localization,~\cite{gangue4} for example, any amount of disorder
drives a system with dimension smaller or equal to two to an insulator
phase. However, this result  was derived considering Anderson localization
effects due to disorder, in the absence of electron-electron interaction.

Several years after the scaling theory of
localization,~\cite{gangue4} the groundbreaking work of
Finkelstein~\cite{finkelstein84} suggested for the first time the
existence of a possible metallic phase in two dimensions for
disordered correlated systems. In Ref.~\onlinecite{finkelstein84}, the 
effects of
correlation and disorder were included by a perturbative
renormalization group approach based on the Fermi liquid theory.
These results are valid in the diffusive regime. Behavior in the
ballistic regime was later considered by Gold and
Dolgopolov,~\cite{gold86}  by Das Sarma and Hwang,
\cite{dassarma99} and  by Zala {\it et
al.}~\cite{zala01}   Most recently, Punnoose and Finkel'stein~\cite{pf05} have
extended the latter's earlier work~\cite{finkelstein84} to give a description
of metallic behavior in two-dimensional semiconductor structures.
However, all these works focus on corrections
to the Fermi liquid theory and as such are appropriate in the
weak coupling regime.~\cite{mott-and} Such
perturbative approaches cannot properly describe strong
correlation effects, such as the approach to the Mott transition,
or the associated transmutation of conduction electrons into local
magnetic moments. These processes are of key importance in many
materials of recent interest, ranging from doped transition metal
oxides and doped semiconductors to organic Mott systems.

To describe systems in which both the correlations and the disorder effects
are strong, complementary methods are required. This is the
situation we examine in the present paper, where we focus on
extensions of the dynamical mean field theory
(DMFT).~\cite{dmftrev,statdmft,tmt}  These extensions are local mean field
approaches, which certainly cannot provide a proper description of
fluctuation effects in low dimensions. Nevertheless, they can be
considered as the simplest approaches incorporating both the Mott
and the Anderson routes to localization. Even simple estimates
indicate that in many systems both of these processes play
comparable roles, and as a result they cannot be considered
separately. Addressing the behavior in this nonperturbative regime
is the main subject of this work.

In the following, we first briefly review the principal physical ideas 
on the interplay of correlations and disorder, as already presented in 
the early works of Mott.~\cite{mott74} We then discuss how this physical 
behavior is naturally described by the extended DMFT approaches which 
are the main subject of this paper.~\cite{statdmft,tmt}

\subsection{Effects of disorder: early ideas}%

\begin{figure}[b]
\begin{center}
\includegraphics[
height=1.3217in, width=3.3723in
]%
{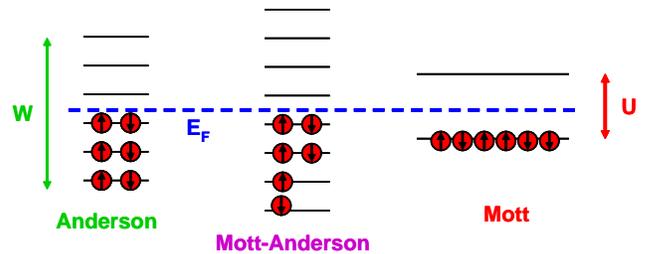} \caption{(Color online) Energy level occupation in the
strongly localized (atomic) limit for an Anderson (left), a Mott
(right), and a Mott-Anderson (center) insulator. In a
Mott-Anderson insulator, the disorder strength $W$ is comparable
to the Coulomb repulsion $U$, and a two-fluid behavior emerges.
Here, a fraction of localized states are doubly occupied or empty
as in an Anderson insulator. Coexisting with those, other states
remain singly occupied forming local magnetic moments, as in a
Mott insulator. Note that the spins of the local moments may be
randomly oriented indicating the absence of magnetic ordering. The
chemical potential is represented by the dashed line.}
\label{fig01}%
\end{center}
\end{figure}

A first glimpse of the basic effect of disorder on the Mott
transition was outlined by Mott,~\cite{mott74} who pointed out
that important consequences of disorder can be seen even when the 
correlations are so strong that the system is in the strongly
localized (atomic) limit. In the absence of disorder, each site
has two energy levels, $\varepsilon_{0}=0$ and
$\varepsilon_{1}=U$, where $U$ is the on-site interaction
potential. If the system is half-filled, then each site is singly
occupied; the levels $\varepsilon_{1}$ remain empty. We have one
local magnetic moment at each site, and a gap equal to $U$ to
charge excitations (as schematically represented in
Fig.~\ref{fig01} on the right).

When disorder is added, each of these energy levels is shifted by a randomly
fluctuating site energy $-W/2<\varepsilon_{i}<W/2$. The situation remains
unchanged for $W<U$, as all the levels $\varepsilon_{1}^{\prime}%
(i)=U+\varepsilon_{i}$ remain empty (for half-filling the chemical potential
is $\mu=E_{F}=U/2$). For larger disorder, those sites with $\varepsilon
_{i}>U/2$ have the level $\varepsilon_{0}^{\prime}(i)=0+\varepsilon_{i}>\mu$
and are empty. Similarly, those sites with $\varepsilon_{i}<-U/2$ have the
excited level $\varepsilon_{1}^{\prime}(i)=U+\varepsilon_{i}<\mu$ and are
doubly occupied. Thus for $W>U$ a fraction of the sites are either doubly
occupied or empty. The Mott gap is now closed, although a fraction of the
sites still remain as localized magnetic moments. We can describe this state as
an inhomogeneous mixture of a Mott and an Anderson insulator (as shown in
the center of Fig.~\ref{fig01}). However, the empty and
doubly occupied sites  have succeeded to completely fill the gap in
the average single particle density-of-states (DOS).

This physical picture of Mott is very transparent and intuitive. The
nontrivial question is how the strongly localized (atomic) limit is approached
as one crosses the metal-insulator transition from the metallic side. To
address this question one needs a more detailed theory for the metal-insulator
transition region, which was not available when the questions posed by
Mott and Anderson were put forward.

\subsection{DMFT description of the Mott-Anderson transition}

The early ideas of Mott have been elaborated and put on a considerably firmer
ground by the development of DMFT approaches.
Here,\cite{dmftrev} the many-body problem is reduced to focusing on a single
site in the lattice  and solving for the dynamics of an electron trying to
delocalize by escaping into its environment. The solution of such an Anderson
impurity problem depends very strongly on the spectrum of electronic states
$\Delta(\omega)$ describing the environment, a quantity determined by an
appropriate self-consistency condition defined by the lattice model in
question.\cite{dmftrev} In the presence of disorder, we have an {\it ensemble}
of single-impurity problems and the self-consistency condition involves the
algebraic average of local quantities over the ensemble. This theory
provides the currently best available description of the Mott metal-insulator
transition, but standard DMFT fails to incorporate Anderson localization
effects.

\ \\ \\
\noindent
{\it Localization effects within DMFT: statDMFT}
\ \\ 

To overcome this difficulty, in a recent work~\cite{statdmft} DMFT was
extended by allowing for spatial fluctuations of the cavity spectrum, now
given by $\Delta_i(\omega)$ (note the index~$i$). This led to the
so-called \textit{stat}DMFT approach, which followed the original method of
Anderson.~\cite{anderson58} Here, one takes the local point of view, but
looks at the typical values of the quantities, that is, to their most
probable values. In particular, one can examine the typical escape rate,
which is given by the typical DOS available
to escape from a given lattice site. These accessible levels are typically
found away from the Fermi energy, reducing the relevant spectral weight
``seen'' by the local electron. The \textit{stat}DMFT approach is intuitively
appealing and physically well-justified; its limitation is the extensive
numerical work required to solve the associated stochastic equations.

\ \\
\noindent
{\it Typical Medium Theory}
\ \\ 

An alternative and much simpler approach was recently provided by the typical
medium theory (TMT),~\cite{tmt} in  which the local spectrum of the
effective medium is replaced by its geometric (typical) average. This 
approach proved able to reproduce many expected features of the Anderson 
localization transition for dimension larger than $2$. It can be
viewed as the conceptually simplest local picture of the Anderson
localization. In this approximation, the spectral weight of the cavity is
reduced as the Anderson transition is approached.

Even before doing any
calculations, one may anticipate that such an elementary manifestation of
incipient localization should dramatically affect the correlation effects when
the electron-electron interaction is turned on. Very recent work by Byczuk
\textit{et al.}~\cite{vollhardt} followed a TMT approach and confirmed that (in
contrast to standard DMFT) it provides a consistent description of both Mott
and Anderson transition in the limiting cases and also of the Mott-Anderson
transition when both interaction and disorder are present. A phase diagram
obtained by employing a numerical renormalization group impurity solver
was presented. However, a more detailed understanding of the physical
behavior in the critical regime, such as possible scaling properties and
power-law behavior, was not elucidated.

To have a better understanding of what is going on, we use a ``4-boson''
impurity solver (in the following call it SB4) of Kotliar and
Ruckenstein~\cite{sb4} to examine the critical behavior at the Mott-Anderson
transition. Before having a complete description of this critical behavior,
it is important to first understand in detail the critical behavior of the
local impurity models, and this is what we concentrate on in this paper.

\subsection{Critical behavior of the impurity models}

it is clear from the above discussion of the atomic limit, as well as from
the previous \textit{stat}DMFT
results,~\cite{statdmft} that for $W>U$ the Mott-Anderson
transition has a qualitatively different character than either the
pure Mott or Anderson transitions. One has to deal with a two-fluid situation,
as a fraction of the electrons become local magnetic moments and the other
fraction remains in magnetically inert configurations.

In this paper, we explore this two-fluid picture from the point of view of
single-impurity problems with different on-site energies. Instead of
addressing the structure of the self-consistent solution, we examine
a simpler problem, where we use a simple model to describe the behavior of the
cavity spectrum. More specifically, we look at the behavior of a collection
of Anderson impurity models, defined by the local site energies
$\varepsilon_{i}$ as, guided by the TMT results,~\cite{ourtmt} we reduce 
the spectral weight of the bath function.
The latter is a measure of the distance $t$ to the transition and we thus
mimic the approach to it.

We find that when the distance $t$ to the transition decreases, depending on
the value of the impurity on-site energy $\varepsilon_{i}$, the quasiparticle
weight $Z_{i}$ goes either to $0$ or $1$. We explore in detail the scaling
behavior of $Z_{i}$ as a function of $t$ and $\varepsilon_{i}$, which can be
expressed in an elegant form using an appropriate $\beta$-function formulation,
as described below. We anticipate that our detailed understanding of this
scaling behavior will allow for an analytical solution of the critical
behavior within the full self-consistent theory. This interesting direction
is left for future work. Note, though, that the analysis presented in this
paper describes properties of quantum impurities in a metallic host and is therefore relevant for that class of problem.

\begin{figure}[b]
\begin{center}
\includegraphics[scale=0.30,angle=-90]
{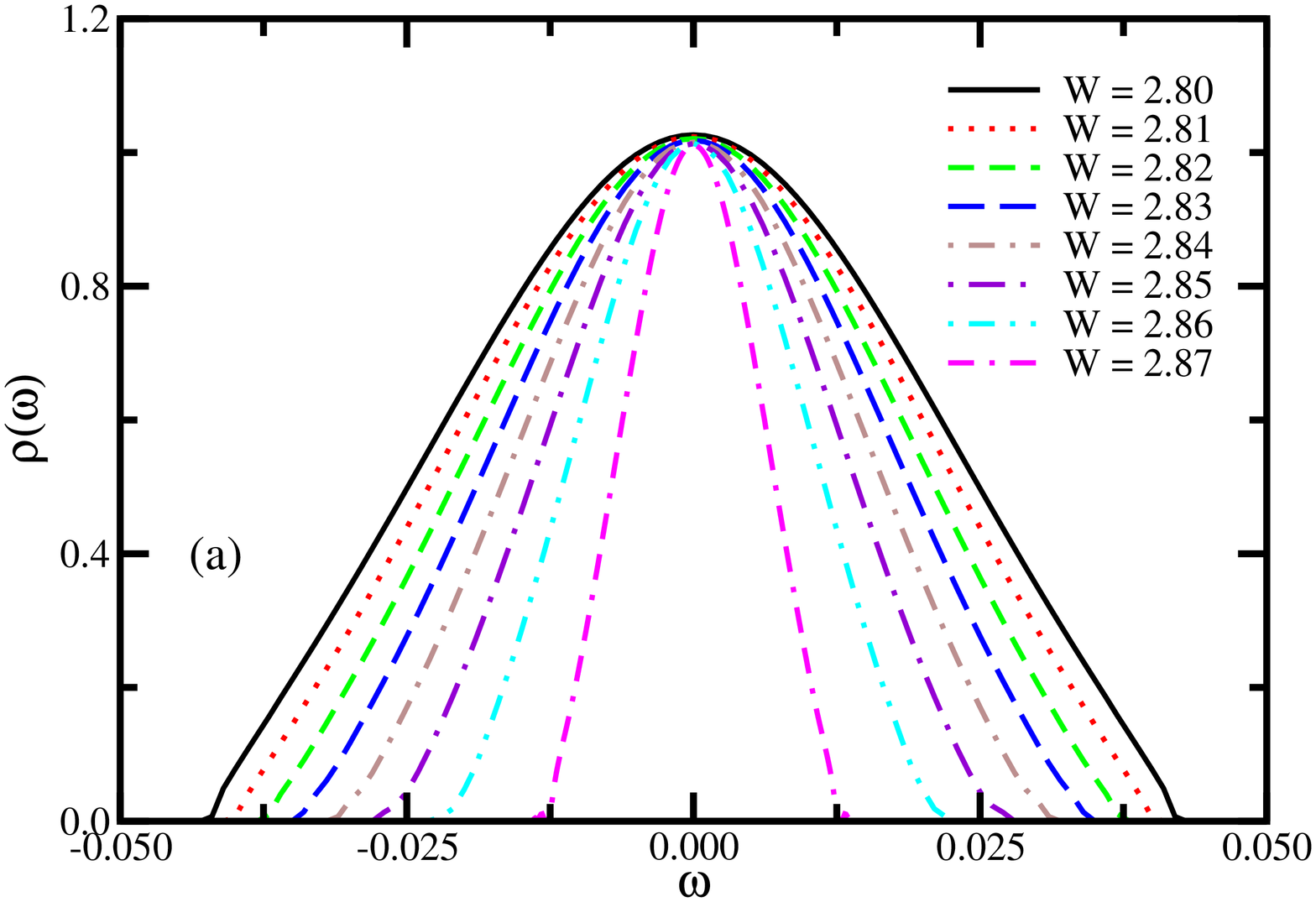}
\includegraphics[scale=0.30,angle=-90]
{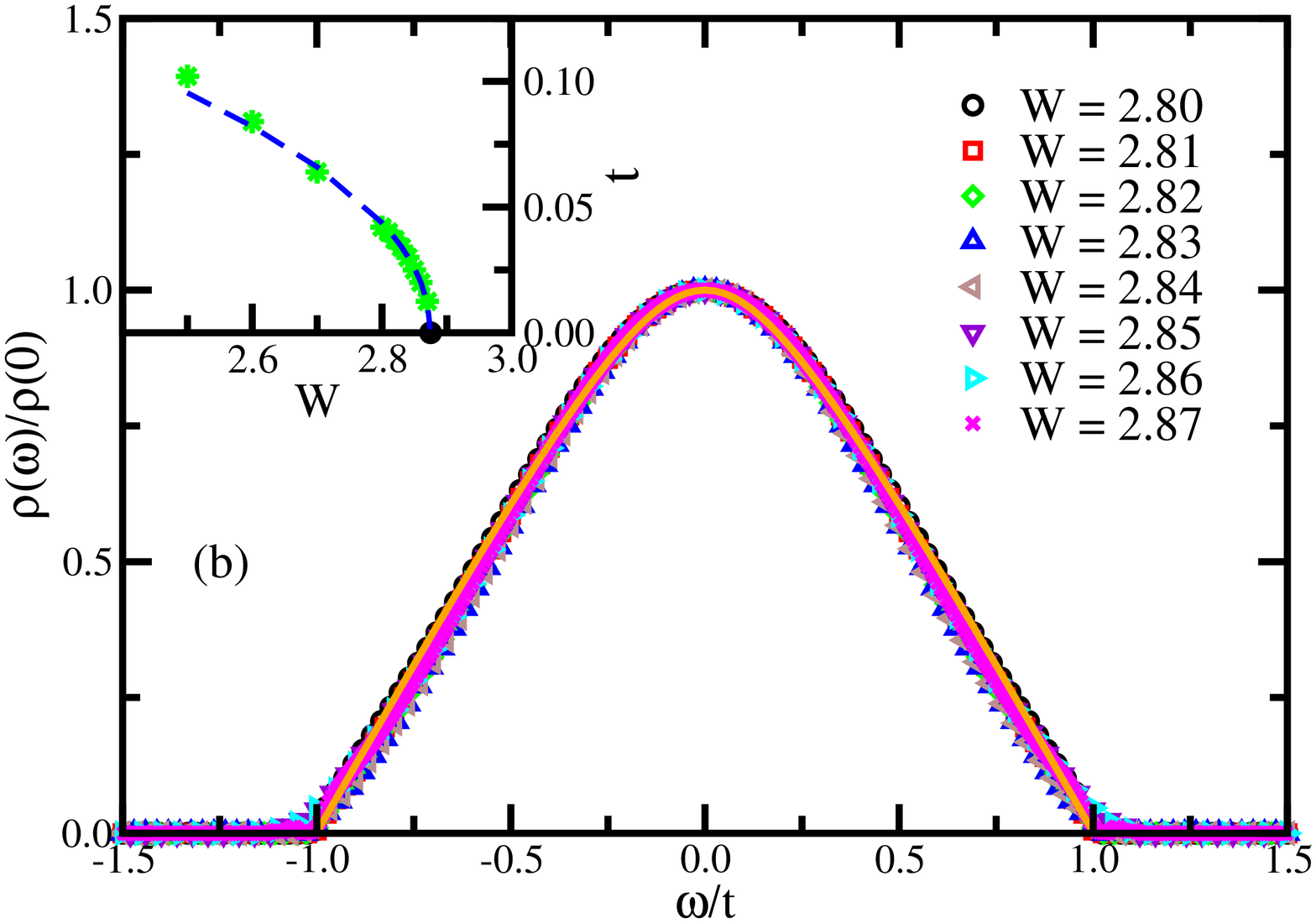}
\end{center}
\caption{(Color online) (a) Evolution of the typical DOS obtained for the
disordered Hubbard model close to the Mott-Anderson transition, approached
by fixing the interaction $U$ and increasing the disorder $W$. The problem
is solved using DMFT-TMT. The disorder is present in the on-site energy,
which follows a uniform distribution of width $W$. The bare DOS is a
semicircle of width $2D=1$.
(b) Scaling of the results in (a). The  typical DOS for different $W$ can be
fit to a single function by scaling the frequency. The inset shows that the
scaling parameter $t$ decreases as the disorder increases and follows a
powerlaw (the dashed line is a powerlaw fitting to the numerical data).}
\label{fig0}
\end{figure}

The paper is organized as follows. In section II, we present the
single-impurity models we solved and the method used to do it. In section
III, we discuss the $\beta$-function formulation valid for the problem in
question. We finish by presenting our conclusions in section IV.

\section{Anderson impurity close to an insulating state}

In DMFT language, a given site is seen as an Anderson impurity that has
the freedom to dynamically choose between a singlet  or a
doublet (local moment) solution upon entering the insulator. The relevant
impurity model is characterized by the on-site repulsion $U$, the on-site
energy $\varepsilon_{i}$ and the total spectral weight $t$ of the cavity
field (this is defined in terms of $t$ below). The hamiltonian describing
the problem is given by
\begin{align}
H &  =\sum_{\vec{k}\sigma}\epsilon_{\vec{k}\sigma}c_{\vec{k}\sigma}^{\dagger
}c_{\vec{k}\sigma}+V\sum_{i\sigma}\left(  c_{i\sigma}^{\dagger}f_{\sigma
}+f_{\sigma}^{\dagger}c_{i\sigma}\right)  \nonumber\\
&  +(\varepsilon_{i}-\mu)\sum_{\sigma}f_{\sigma}^{\dagger}f_{\sigma
}+Uf_{\uparrow}^{\dagger}f_{\uparrow}f_{\downarrow}^{\dagger}f_{\downarrow
},\label{1imp}%
\end{align}
where $c_{\vec{k}\sigma}$ and $f_{\sigma}$ are annihilation operators of
extended and localized electrons, respectively, and $V$ and $\epsilon_{\vec
{k}}$ define the cavity field $\Delta(\omega_{n})=V^{2}\sum_{\vec{k}%
}[1/(\omega_{n}-\epsilon_{\vec{k}})]$. Alternatively, the cavity function can
be written in the spectral form
\begin{equation}
\Delta(\omega_{n})=V^{2}\int_{-\infty}^{\infty}d\omega\frac{\rho_{typ}(\omega
)}{\omega_n-\omega}.\label{delta}%
\end{equation}

We used two different forms for $\rho_{typ}(\omega)$:

\begin{enumerate}
\item The typical DOS calculated self-consistently
when approaching the Mott-Anderson transition.\cite{ourtmt} An example
can be seen in Fig.~\ref{fig0} for $U=1.75$ and different values of the
breadth of the disorder $W$. We
find that, as the transition is approached, the spectrum retains the same
shape, while mainly the total spectral weight, determined here by the
spectrum width, decreases [see panel (a)].\cite{note}
Besides, as shown in panel (b),
the results can be scaled to a single function, with the scaling parameter
$t$ (inset) having a powerlaw dependence on $W$. Here $t$ is a measure of
the spectrum width.

\item A featureless model bath given by $1$ for $-t/2<\omega<t/2$ and zero
otherwise, with $t\rightarrow0$ at the transition.
\end{enumerate}

As we shall see, the
qualitative critical
behaviors of the impurity models are not affected by the specific choice of the
spectral function, as identical critical exponents are obtained for the two
models. This is important, since a simple featureless model offers a
considerable advantage for analytical work.

The impurity model of Eq.~(\ref{1imp}) was solved at zero temperature using
the SB4 method,~\cite{sb4} which provides a parametrization of the low-energy
(quasiparticle) part of the local Green's function, given by

\begin{equation}
G_i(\omega_n)=\frac{Z_i}{i\omega_n-\tilde{\varepsilon}_i-Z_i\Delta
(\omega_n)},
\end{equation}
where $Z_i$ is the local quasiparticle weight and $\tilde{\varepsilon}_i$
is the renormalized site energy. The approach consists of determining the
site-dependent parameters $e_{i}$, $d_{i}$ and $\tilde{\varepsilon}_{i}$ by
the following equations
\begin{equation}
-\frac{\partial Z_{i}}{\partial e_{i}} \frac{1}{\beta} \sum_{\omega_{n}}
\Delta(\omega_{n}) G_{i}(\omega_{n})=Z_{i} \left(  \mu+\tilde{\varepsilon}_{i}
-\varepsilon_{i} \right)  e_{i},
\end{equation}

\begin{equation}
-\frac{\partial Z_{i}}{\partial d_{i}} \frac{1}{\beta} \sum_{\omega_{n}}
\Delta(\omega_{n}) G_{i}(\omega_{n})=Z_{i} \left(  U-\mu-\tilde{\varepsilon
}_{i} +\varepsilon_{i} \right)  d_{i},
\end{equation}

\begin{equation}
\frac{1}{\beta} \sum_{\omega_{n}} G_{i}(\omega_{n})= \frac{1}{2} Z_{i} \left(
1-e_{i}^{2}+d_{i}^{2} \right)  ,
\end{equation}
where $Z_{i}=2(e_{i}+d_{i})^{2}[1-(e_{i}^{2}+d_{i}^{2})]/ [1-(e_{i}^{2}
-d_{i}^{2})^{2}]$ in terms of $e_i$ and $d_i$ and $\mu=U/2$.

\begin{figure}[t]
\begin{center}
\includegraphics[scale=0.30,angle=-90]
{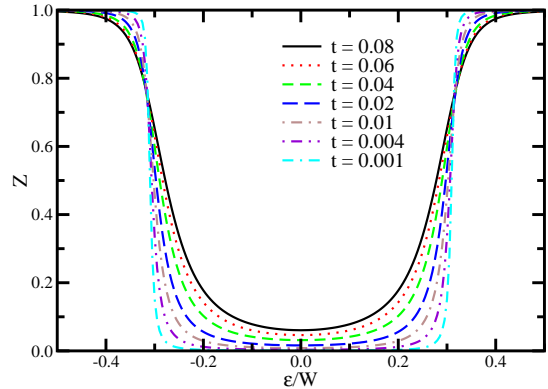}
\end{center}
\caption{(Color online) Quasiparticle weight $Z$ as a function of the on-site
energy $\varepsilon$ for the single-impurity problems close to the
Mott-Anderson transition. A flat bath (see text) was used. The transition is
approached when the bath weight $t$ decreases towards $0$. The parameters
used were $U=1.75$ and $W=2.8$.}%
\label{fig1}%
\end{figure}

The results for the quasiparticle weight $Z_i$ as a function of $\varepsilon
_{i}$ are shown in Fig.~\ref{fig1} for the flat bath function explained
above. Almost identical results are found for the self consistent typical
bath of Fig.~\ref{fig0}. We shall examine the situation where the spectral
weight $t$ goes to zero, meaning that the system goes towards the
Mott-Anderson transition.

\begin{figure}[b]
\begin{center}
\includegraphics[scale=0.30,angle=-90]
{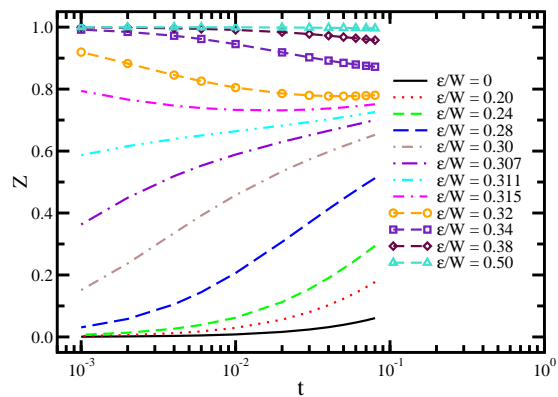}
\end{center}
\caption{(Color online) Same results as in Fig.~\ref{fig1} for the
quasiparticle weight $Z$ plotted in a different way: as a function of $t$ for
different values of $\epsilon/W$ (instead of as a function of $\epsilon/W$
for different values of $t$, as before). We present the results only for
positive site energies, as a similar behavior holds for negative ones. Other
parameters as in Fig.~\ref{fig1}.}%
\label{fig8}%
\end{figure}

Considering many single-impurity problems, we observe a two-fluid
picture, just as in the limit earlier analyzed by Mott.~\cite{mott74} Indeed,
these results correspond to the same atomic limit discussed by Mott, since,
although the hopping itself is still finite, the cavity field ``seen'' by
the impurities goes to zero in the current case. One can either look at the
results as a function of $\varepsilon_i$ for different values of $t$, as
presented in Fig.~\ref{fig1}, or look at the same data plotted as a function
of $t$ for different values of $\varepsilon_i$, as shown in Fig.~\ref{fig8}.

As in the atomic limit, the sites with $|\varepsilon_{i}|<U/2$ turn into
local moments and have vanishing quasiparticle weight $Z_i\rightarrow0$.
The remaining sites show $Z_i\rightarrow1$, as they are either doubly occupied,
which corresponds to those with $\varepsilon_{i}<-U/2$, or
empty, which is the case for those sites with $\varepsilon_{i}>U/2$.
Consequently, as the transition is approached, the curves $Z(\varepsilon_{i},
t)$ ``diverge'' and approach either $Z=0$ or $Z=1$. These values of $Z$ can
thus be identified as two stable fixed points for the problem in question,
as we discuss below.

Note that in Fig.~\ref{fig8} we restrict the results to positive energy
values, as a similar behavior is observed for negative $\varepsilon_i$. In
this case, there is precisely one value of the site energy $\varepsilon_{i}
=\varepsilon^{\ast}$, for which $Z(\varepsilon^{\ast},t)\rightarrow
Z^{\ast}$. This corresponds to the value of $\varepsilon_{i}$ below which
$Z$ ``flows'' to $0$ and above which $Z$ ``flows'' to $1$. In other words,
it corresponds to an unstable fixed point. Just as in the atomic
limit, $\varepsilon^{\ast}$ is equal to $U/2$ ($\varepsilon^{\ast}/W=0.3125$
in Fig.~\ref{fig8}, where $U=1.75$ and $W=2.8$).

Interestingly, the family of curves in Fig.~\ref{fig8} looks similar to those
seen in some other
examples of quantum critical phenomena. In fact, one can say that the
crossover scale $t$ plays the
role of the reduced temperature, and the reduced site energy $\delta
\varepsilon=\left(  \varepsilon_{i}-\varepsilon^{\ast}\right)  /\varepsilon
^{\ast}$ that of the control parameter of the quantum critical point. As the
site energy is tuned at $t=0$, the impurity model undergoes a phase transition
from a singlet to a doublet ground state. Quantum fluctuations associated with
the metallic host introduce a cutoff and round this phase transition, which
becomes sharp only in the $t\rightarrow0$ limit. In the following, we develop
a detailed scaling analysis of the $t=0$, $\delta\varepsilon=0$ fixed point
based on our numerical and analytical results
obtained from the SB4 impurity model solution.~\cite{sb4}

\section{$\beta$-function formulation}

\begin{figure}[t]
\begin{center}
\includegraphics[scale=0.30,angle=-90]
{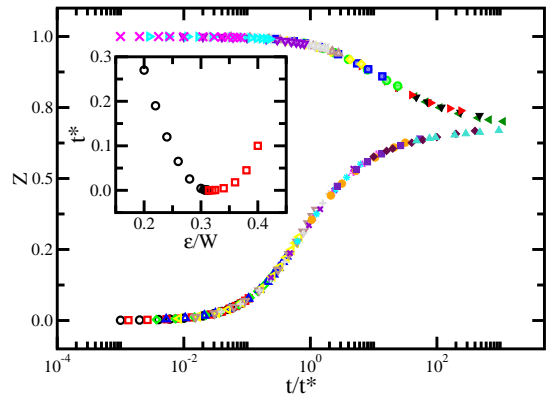}
\end{center}
\caption{(Color online) Quasiparticle weight $Z$ as a function of $t/t^{\ast
}(\delta\varepsilon)$ showing that the results for different $\varepsilon$ can
be collapsed onto a single scaling function with two branches. The results for
different $\varepsilon$ correspond to different symbols. The inset shows the
scaling parameter $t^{\ast}$ as a function of $\varepsilon/W$ for the upper
(squares) and bottom (circles) branches. Other parameters as in
Fig.~\ref{fig1}.}%
\label{fig2}%
\end{figure}

\begin{figure}[b]
\begin{center}
\includegraphics[scale=0.30,angle=-90]
{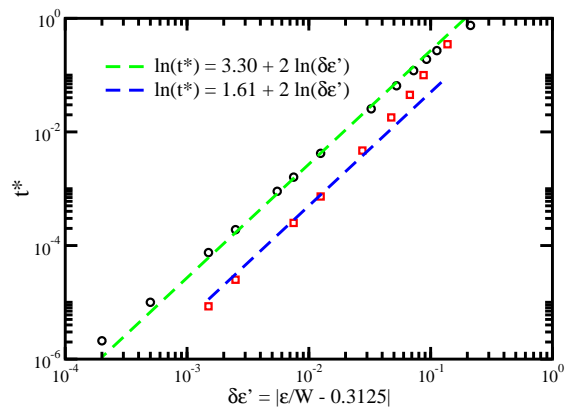}
\end{center}
\caption{(Color online) Scaling parameter $t^{\ast}$ as a function of
$\delta\varepsilon$ for the scaling procedure presented in Fig.~\ref{fig2}.
The results for the upper ($Z\rightarrow1$) and the bottom ($Z\rightarrow0$)
branches correspond to squares and circles, respectively. For both branches,
$t^{\ast}$ follows a powerlaw with exponent equal to $2$.}%
\label{fig3}%
\end{figure}

Our numerical solutions provide evidence that as a function of $t$ the
``charge'' $Z(t)$ ``flows'' away from the unstable ``fixed point'' $Z^{\ast}$,
and towards either stable ``fixed points'' $Z=0$ or $Z=1$. The structure of
these flows show power-law scaling as the scale $t\rightarrow0$;
this suggests that it should be possible to collapse the entire family of
curves $Z(t,\delta\varepsilon)$ onto a single universal scaling function
\begin{equation}
Z(t,\delta\varepsilon)=f[t/t^{\ast}(\delta\varepsilon)], \label{scaling}
\end{equation}
where the crossover scale $t^{\ast}(\delta\varepsilon)= C^{\pm}|\delta
\varepsilon|^{\phi}$ around the unstable fixed point. Remarkably, we have been
able to scale the numerical data precisely in this fashion, see
Fig.~\ref{fig2}, and extract the form of $t^{\ast}(\delta\varepsilon)$.
According to Fig.~\ref{fig3}, $t^{\ast}(\delta\varepsilon)$ does vanish in a
power law fashion at $\delta\varepsilon=0$, with exponent $\phi=2$ and the
amplitudes  $C^{\pm}$ differ by a factor close to two for $Z \gtrless Z^*$.

\begin{figure}[b]
\begin{center}
\includegraphics[scale=0.25,angle=-90]
{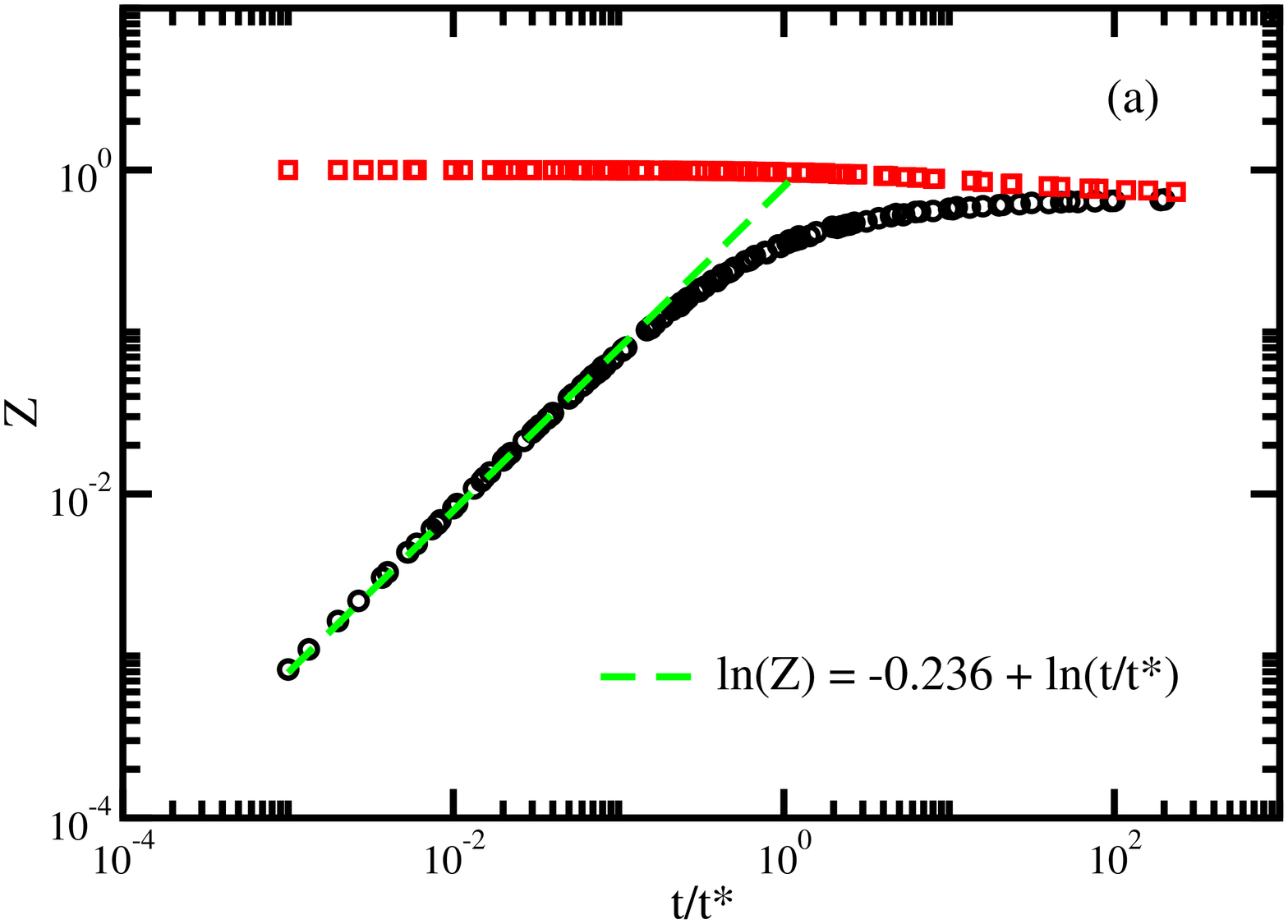}
\includegraphics[scale=0.25,angle=-90]
{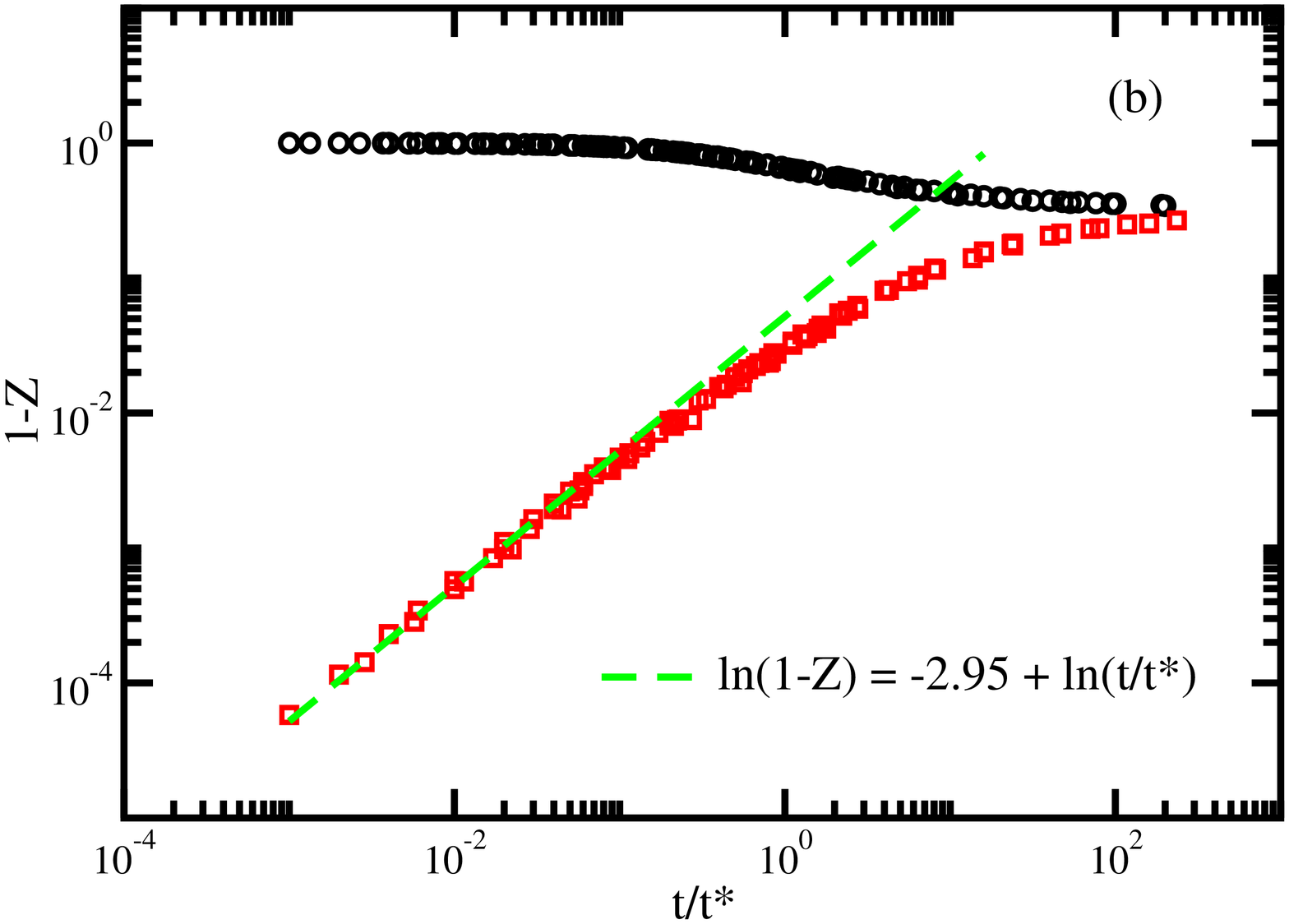}
\includegraphics[scale=0.25,angle=-90]
{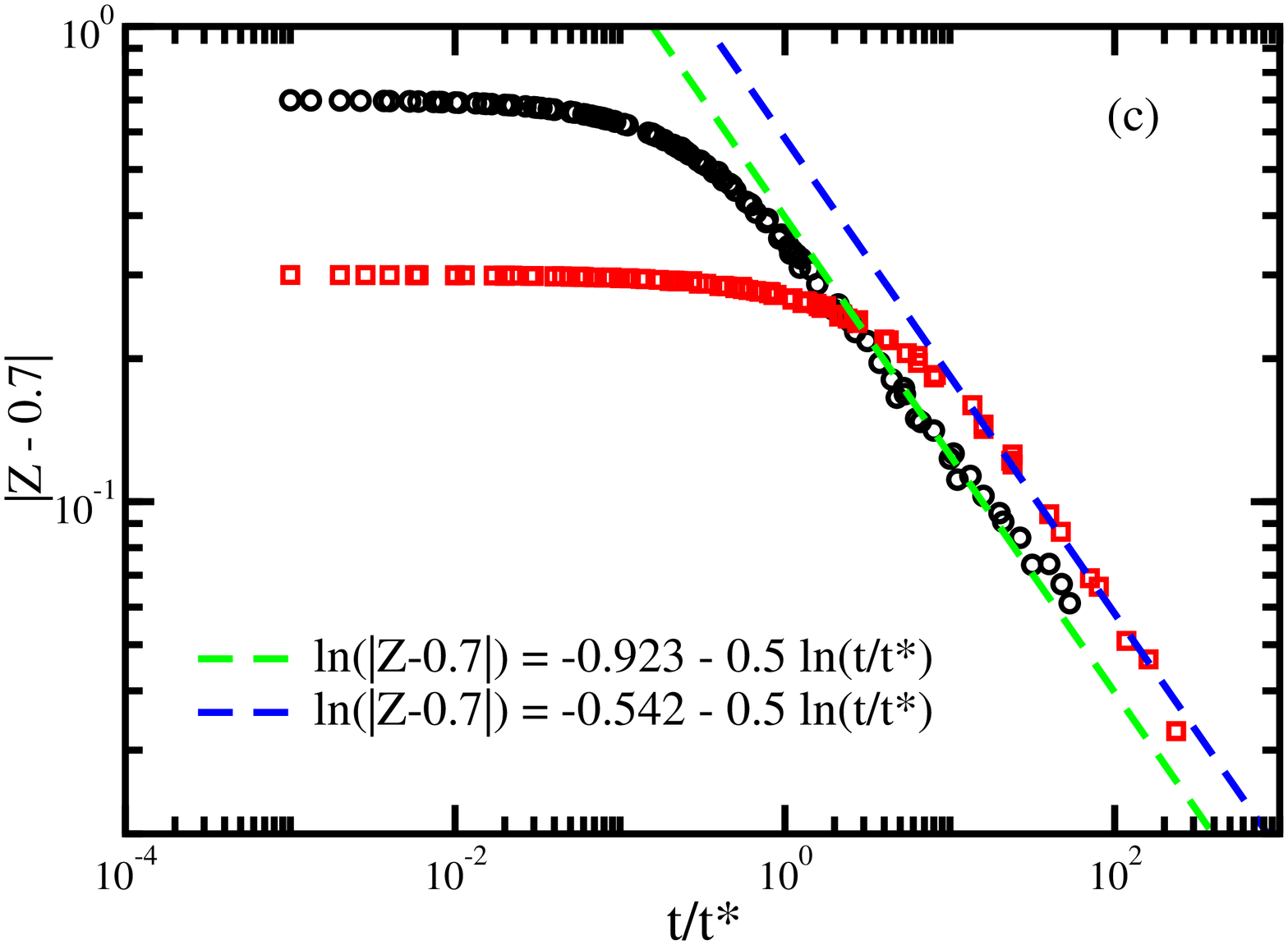}
\end{center}
\caption{(Color online) Same numerical data presented in Fig.~\ref{fig2} now
showing the powerlaw behavior in three different cases: (a) $Z(t)\sim t$ for
the sites with $\varepsilon<U/2$ (bottom branch) in the limit $t
\rightarrow0$, (b) $Z(t)\sim1-B^+t$ for the sites with $\varepsilon>U/2$
(upper branch) in the limit $t \rightarrow0$ and (c)~$|Z(t)-Z^{\ast}| \sim
A^{\pm}t^{-1/2}$ for the two branches when $t>>1$. In all panels, squares and
circles correspond to the upper and the bottom branches, respectively.}%
\label{fig4}%
\end{figure}

As shown in Fig.~\ref{fig2}, the scaling function $f(x)$ where $x=t/t^{\ast}
(\delta\varepsilon)$ presents two branches: one for $\varepsilon_i<
\varepsilon^{\ast}$ and other for $\varepsilon_i>\varepsilon^{\ast}$. We
found that for $x \rightarrow0$ both branches of $f(x)$ are linear in $x$
[Fig.~\ref{fig4}(a) and \ref{fig4}(b)], while for $x\gg1$ they merge, i.e.
$f(x)\rightarrow Z^{\ast}\pm A^{\pm}x^{-1/2}$ [Fig.~\ref{fig4}(c)]. As can
be seen in the first two panels, in the limit $t \rightarrow0$, the
curve corresponding to $\varepsilon_i<U/2$ has $Z(t)=B^- t$, while that for
$\varepsilon_i>U/2$ follows $1-Z(t)=B^+t$. These
results are for a flat cavity field but, as mentioned earlier, we checked
that the same exponents are found also for other bath functions, meaning that
they are independent of the exact form of the cavity field.

The power-law behavior and the respective exponents observed numerically
in the three limits above are confirmed if we solve the SB
equations analytically close to the transition ($t \rightarrow 0$). For
small $\varepsilon_i$, we have that $Z_i$ goes as

\begin{equation}
Z_i \approx \frac{64 V^2}{U^2} t \left( 1+\frac{8}{U^2} \varepsilon_i^2
\right), \label{lim1}
\end{equation}
that is, has a linear dependence on $t$, which corresponds to the same
exponent obtained in the fitting shown in Fig.~\ref{fig4}(a). In the opposite
limit of large $\varepsilon_i$ (more specifically, $\varepsilon_i>>U/2$ and
$\tilde{\varepsilon_i}^2>>t$), in accordance with what is seen in
Fig.~\ref{fig4}(b), $Z_i$ differs from $1$ by a term which is linear in $t$,

\begin{equation}
1-Z_i \approx 3\times10^{-6}\frac{V^2}{\varepsilon_i^2}t. \label{lim2}
\end{equation}
Finally, for $\varepsilon_i$ around and close to $\varepsilon^{\ast}=U/2$,
we have

\begin{equation}
Z_i \approx \frac{2}{3}+\frac{1}{9V}\sqrt{\frac{2}{t}}
\left( \varepsilon_i-U/2 \right), \label{lim3}
\end{equation}
once more confirming the exponent equal to $-0.5$ seen in the numerical
results of Fig.~\ref{fig4}(c).

The results in eqs.~(\ref{lim1}-\ref{lim3}) are valid for a simple flat bath.
In case of a more general form for $\rho_{typ}(\omega)$, assuming
$t \rightarrow 0$, the following substitution should be made in the above
equations

\begin{equation}
t \rightarrow \int_{-t/2}^{t/2} d\omega \rho_{typ}(\omega). \label{intrho}
\end{equation}
Considering that the scaling properties shown in Fig.~\ref{fig0}(b) hold,
that is,

\begin{equation}
\rho_{typ}(\omega)/\rho_{typ}(0)=g(\omega/t),
\end{equation}
where $g(y=\omega/t)$ is the scaling function, we can rewrite
Eq.~(\ref{intrho}) as

\begin{equation}
t \rightarrow t \left[ \rho_{typ}(0) \int_{-1/2}^{1/2} dy g(y) \right] \sim t.
\end{equation}
This means that the same exponents for the power-law behavior in $t$ are
found independent of the specific form of the cavity field, as pointed out
before.

In the following, we rationalize these findings by defining an appropriate
$\beta$-function which describes all the fixed points and the corresponding
crossover behavior. Let us assume that
\begin{equation}
\frac{dZ(t,\delta\varepsilon)}{d\ln t}=-\beta(Z) \label{defbeta}
\end{equation}
is an explicit function of $Z$ only, but not of the parameters $t$ or
$\delta\varepsilon$. The desired structure of the flows would be obtained if
the $\beta$-function had three zeros: at $Z=0$ and $Z=1$ with negative
slope (stable fixed points) and one at $Z=Z^{\ast}$ with positive slope
(unstable fixed point).

\begin{figure}[ptb]
\begin{center}
\includegraphics[scale=0.30,angle=-90]
{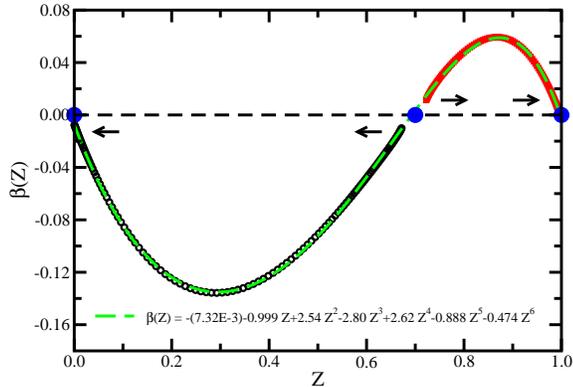}
\end{center}
\caption{(Color online) $\beta$-function obtained as described in the text for
the Anderson impurity models close to the Mott-Anderson transition. The filled
circles indicate the three fixed points found for this problem. The arrows
indicate how $Z$ flows \textit{to} the stable points ($Z=0$ and $Z=1$) and
\textit{from} the unstable one ($Z\approx0.7$).} %
\label{fig6}%
\end{figure}

\subsection{How to determine the $\beta$-function?}

Assuming that the general structure of these flows can be described in a
$\beta$-function language similar to that used in the context of a
renormalization group approach, we outline the procedure to obtain $\beta(Z)$
from the numerical data.

The integration of Eq.~(\ref{defbeta}) can be written in the form of
Eq.~(\ref{scaling}) as%
\begin{equation}
Z=f[t/t^{\ast}(Z_{o})],
\end{equation}
where $Z_{o}$ is the initial condition (a function of $\delta\varepsilon$).
With $x=t/t^* $ as before, Eq.~(\ref{scaling}) can be rewritten as
\begin{equation}
\beta(Z)=-xf^{\prime}(x). \label{beta}
\end{equation}

The numerical data for $Z=f(x)$ as a function of $x$ is presented in
Fig.~\ref{fig2}. It is possible to fit the two branches in this figure with
simple functions, as the results in Fig.~\ref{fig4} suggest. Thus, using
eq.~(\ref{beta}), the $\beta$-function in terms of $x(Z)$ is determined,
which can finally be rewritten in terms of $Z$. Carrying
out this procedure, we obtain $\beta(Z)$ as shown in Fig.~\ref{fig6}. In
accordance with what was discussed above, we see that $\beta(Z)$ has three
fixed points, as indicated in the figure by filled circles. $Z=0$ and $Z=1$
are stable, while $Z \approx 0.7$ is the unstable fixed point.

This procedure proves that a $\beta$-function formulation is valid in the case
of the Anderson impurity models when approaching the Mott-Anderson transition.
The scaling behavior and the associated $\beta$-function observed here reflect
the fact these impurity models have two phases (singlet and doublet) when
entering the insulator. The two stable fixed points describe these two phases,
while the unstable fixed point $Z^{\ast}$ describes the phase transition,
which is reached by tuning the site energy. To completely solve the problem
of the Mott-Anderson transition, however, the bath function ``seen'' by the
{\it ensemble} of single-impurity problems has to be self-consistently
determined. The results obtained in this case are left to be discussed
elsewhere.~\cite{ourtmt}

\section{Conclusions}

In this paper, we focused on understanding the behavior of  Anderson
impurity problems in a model bath chosen to mimic the approach to the
Mott-Anderson transition. We presented numerical and analytical results
that portray a simple scaling behavior in the critical regime where the bath
spectral weight becomes vanishingly small. Our scaling analysis clarifies the
emergence of the two-fluid behavior at the critical point, which can be
considered as a first step towards constructing a full self-consistent
description of the Mott-Anderson transition.

The scaling behavior and the associated $\beta$-function observed by us
reflect the fact that, when entering the insulator, the impurity models
have two phases (singlet and doublet), undergoing a phase transition as
the site energy is tuned. In this
sense, most of our conclusions represent properties of the Anderson impurity
models in the relevant strong-coupling limit, and as such have a potential
use beyond the possible application to a specific
self-consistent scheme such as the TMT approach.\cite{tmt} Indeed, we
expect that similar critical scaling properties apply in a more sophisticated
\textit{stat}DMFT theory,\cite{statdmft} where the impurity is embedded in
a statistical {\it ensemble} of fluctuating baths. Applied to this problem,
the present work opens an avenue to analytically characterize the scaling of
the relevant distribution functions in the critical regime.

Incorporating what we learned about the impurity models into these
self-consistent schemes is an interesting and important problem left for
future work.

This work was supported by NSF grants DMR-0234215 (VD)  
and DMR-0528969 (GK). MCOA was supported in part by NSF grant DMR-0312495.


\begin{thebibliography}{9}
\bibitem {note} Within TMT, the typical DOS represents the extended states
of the system and as such has its spectral weight going to zero as the
Mott-Anderson is approached. The (arithmetic) average DOS includes both
extended and localized states and it is its spectral weight that is
normalized within this method.
\end{thebibliography}
\end{document}